\documentclass{aastex}          
\usepackage{spr-astr-addons}    
\usepackage{epstopdf}
\usepackage{url}

\begin{document}
%
\title{Footprints of the weak s-process in the
carbon-enhanced metal-poor star ET0097}

\shorttitle{Footprints of the weak s-process in ET0097}
\shortauthors{Yang et al.}

\author{Guochao Yang\altaffilmark{1}, Hongjie Li\altaffilmark{1,2}, Nian Liu\altaffilmark{3},
Wenyuan Cui\altaffilmark{1}, Yanchun Liang\altaffilmark{4} and Bo Zhang\altaffilmark{1,*}}

\altaffiltext{1}{Department of Physics, Hebei Normal University,
NO. 20 Road East of 2nd Ring South, Shijiazhuang, 050024, China}
\altaffiltext{2}{School of Sciences, Hebei University of Science and Technology,
Shijiazhuang, 050018, China}
\altaffiltext{3}{Astronomy Department, Beijing Normal University,
No. 19, XinJieKouWai St., HaiDian District, Beijing, 100875, China}
\altaffiltext{4}{Key Laboratory of Optical Astronomy, National Astronomical Observatories,
Chinese Academy of Sciences, 20A Datun Road, Chaoyang District, Beijing, 100012, China}
\altaffiltext{*}{E-mail address: zhangbo@mail.hebtu.edu.cn}

\begin{abstract}
Historically, the weak s-process contribution to metal-poor stars
is thought to be extremely small, due to the effect of the
secondary-like nature of the neutron source $^{22}$Ne($\alpha,n$)$^{25}$Mg
in massive stars, which means that metal-poor ``weak s-process
stars'' could not be found. ET0097 is the first observed
carbon-enhanced metal-poor (CEMP) star in
the Sculptor dwarf spheroidal galaxy. Because C is enriched and the
elements heavier than Ba are not overabundant, ET0097 can be
classified as a CEMP-no star. However, this star shows overabundances
of lighter n-capture elements (i.e., Sr, Y and Zr). In this
work, having adopted the abundance decomposition approach, we investigate
the astrophysical origins of the elements in ET0097. We find that
the light elements and iron-peak elements (from O to Zn) of the star
mainly originate from the primary process of massive stars and
the heavier n-capture elements (heavier than Ba) mainly come from
the main r-process. However, the lighter n-capture elements such as Sr, Y
and Zr should mainly come from the primary weak s-process. The
contributed fractions of the primary weak s-process to the Sr, Y and Zr
abundances of ET0097 are about 82\%, 84\% and 58\% respectively, suggesting
that the CEMP star ET0097 should have the footprints of the weak s-process.
The derived result should be a significant evidence that the weak
s-process elements can be produced in metal-poor massive stars.
\end{abstract}


%
\section{Introduction}

Stellar chemical abundances reflect the accumulated results of
various nucleosynthesis processes. Study of the stellar chemical
abundance patterns is a vital task for understanding the
different nucleosynthesis processes \citep{sneden2008}. The light
elements and iron-peak (Fe-peak) elements can be produced through
charged-particle fusion reactions, whereas the heavier elements are
produced through the neutron-capture (n-capture) process which
includes the slow n-capture process (s-process) and the rapid
n-capture process (r-process) \citep{burbidge1957}. The s-process
is further divided into two sub-processes: the main s-process and
the weak s-process. The main s-process occurs in thermally
pulsing asymptotic giant branch (TP-AGB) stars with low- and
intermediate-mass and produces n-capture elements, especially
the heavier n-capture elements with $Z \geq 56$
\citep{gallino1998,busso1999,busso2001}. The helium-burning (He-burning)
cores and the carbon-burning (C-burning) shells of massive stars are
the sites of the weak s-process which mainly produce the lighter
n-capture elements with $38 \leq Z \leq 47$
\citep{lamb1977,raiteri1991,raiteri1992,raiteri1993}.

On the other hand, the r-process can also be classified into two
sub-processes: the main r-process and the weak r-process.
Although the r-process nucleosynthesis mechanisms have
been investigated by many pioneering works
\citep[e.g.,][]{burbidge1957,hillebrandt1978,mathews1990,woosley1994,wheeler1998},
different works gave different suggestions \citep{sneden2008}.
However, two possible sites for the r-process
are commonly proposed.
The first possible site relies on SNe II \citep{woosley1994,qian1996}.
\citet{travaglio1999} suggested that the main r-process should
occur in SNe II with an initial mass range of $8-10\,M_\odot$
and dominantly generate heavier n-capture elements. While,
\citet{qian2007} proposed that the weak r-process should occur in
SNe II with the initial mass range of $11-25\,M_\odot$ and be a
source for lighter n-capture elements. Furthermore,
based on the abundance analysis of metal-poor stars,
\citet{montes2007} concluded that the ``lighter element primary
process (LEPP)'' (or weak r-process) produces a uniform and unique
abundance pattern. The second possible site refers to the mergers
of binary neutron stars (NS-NS) or neutron star-black hole (NS-BH)
systems \citep{lattimer1974,lattimer1976,eichler1989}.
\citet{freiburghaus1999} suggested that the neutron star mergers
(NSMs) should be
responsible for the heavier r-process elements A $\gtrsim$ 130.
\citet{korobkin2012} further suggested that compact
binary mergers (CBMs) can produce a robust r-process abundance
pattern. Moreover, \citet{matteucci2014} proposed that the theoretical
production of the CBMs alone could entirely explain the observed
Eu abundance in the Galaxy.
Recently, \citet{wanajo2014} and \citet{just2015} described that
the NSMs can successfully generate not only heavy r-process
elements but also light r-process elements. Although each of
the two suggested sites above provides an important clue
for understanding the nucleosynthetic mechanisms of the r-process,
the actual astrophysical sites of the r-process remain under debate
\citep{ishimaru2015,goriely2016}.

Carbon-enhanced metal-poor (CEMP) stars are an interesting
stellar class which is composed of stars with low metallicities
and overabundant C. Based on the abundance
patterns of n-capture elements, CEMP stars can be further
divided into four sub-classes \citep{beers2005}:
(1) CEMP-r stars: [Eu/Fe] $>$ 1.0.
Stars of this sub-class are rare. CS22892-052 has been
classified as an example of CEMP-r star and shows no sign
of binarity \citep{sneden2003b,hansen2011}.
(2) CEMP-s stars: [Ba/Fe] $>$ 1.0 \& [Ba/Eu] $>$ 0.5.
These stars are suggested to be in binary systems.
The overabundant s-process elements of one CEMP-s star are thought to
originate from its former-AGB companion \citep{lucatello2005}.
(3) CEMP-r/s stars: 0.0 $<$ [Ba/Eu] $<$ 0.5.
The astrophysical origins of CEMP-r/s stars are in debate,
as these stars show overabundances of both s-process and
r-process elements.
\citet{jonsell2006} summarized 9 possible formation
scenarios to explain the peculiar abundance pattern. They found
that each of the scenarios alone was not enough to explain the enrichment
of both s-process and r-process elements.
(4) CEMP-no stars: [Ba/Fe] $<$ 0.0.
The origins of CEMP-no stars have been studied by many works
\citep[e.g.,][]{bromm2003,ryan2005,frebel2007,masseron2010,gilmore2013,norris2013}.
Based on the radial velocity analysis for the CEMP-no stars,
\citet{starkenburg2014} found that only two of the 15 CEMP-no sample
stars show signatures of binarity.
On the other hand, relied on the abundance analysis of the bright
CEMP-no star BD+44$^\circ$493, \citet{placco2014} and \citet{roederer2016}
suggested that this CEMP-no star could well be a second-generation star.
Recently, with much improved statistics, \citet{hansen2016} concluded that
CEMP-no stars may indeed be bona-fide second-generation stars,
formed from natal gas clouds polluted by the very first (massive) stars.
Furthermore, they also suggested that these stars are not consistent with any
mass-transfer mechanism.

ET0097 ([Fe/H] = $-$2.03) is the first observed CEMP star in the
Sculptor dwarf spheroidal galaxy. \citet{skuladottir2015} found that
this star can be classified as a CEMP-no star, since the C abundance
is high and the elements heavier than Ba are not overabundant.
For lighter n-capture elements (i.e., Sr, Y and Zr),
the abundances of this star are higher than the average abundances
of other stars in Sculptor and C-normal stars in the Galactic
halo. They proposed that, (1) the reason for C enhancement may be that
the gas cloud in which ET0097 formed contains the material from faint SNe;
(2) the lighter n-capture elements should come from the weak r-process
or the weak s-process, while there were not enough evidences to
determine which process is the main contributor. Because ET0097 is
the first observed CEMP-no star in Sculptor, it is important to
investigate the astrophysical origins of its elements, especially
the lighter n-capture elements Sr, Y and Zr.

In this work, we adopt the abundance decomposition approach presented
by \citet{li2013a} to investigate the astrophysical origins of
the elements of ET0097. The abundance decomposition approach of the
sample star is described in \S\,2. The results and
discussions are provided in \S\,3. The conclusions are presented in
\S\,4.

\section{Abundance decomposition approach of the sample star}

The metal-poor stars CS 22892-052 and CS 31082-001 are deemed as the
prototypes of ``main r-process stars'', as their abundance pattern of
heavier n-capture elements matches the r-process abundance pattern
of the solar system closely
\citep{cowan1991,truran2002,wanajo2006,sneden2008}, whereas their
abundances of lighter n-capture elements are deficient relative to
the r-process abundances of the solar system. On the other hand, the
metal-poor stars HD 122563 and HD 88609 are called as ``weak
r-process stars'', because their abundances of lighter n-capture
elements are excessive and their abundances of heavier n-capture
elements are deficient \citep{westin2000,johnson2002,honda2004}.
Based on the
abundances of the main r-process stars and the weak r-process stars,
\citet{li2013a} derived the abundances of the main r- and weak
r-process with iterative approach and found that almost all the
metal-poor stars have been polluted by both main r- and weak
r-process material.

The primary light elements and primary Fe-peak elements are generated
in the massive stars ($M > 10\,M_\odot$) through charged-particle
reactions \citep{heger2010}. The ratios [Sr/Fe] $\simeq$ 0
in weak r-process stars mean that weak r-process elements
and Fe-peak elements are ejected together from the massive stars.
\citet{li2013b} combined the primary-like abundances (i.e., the
yields have no correlation with the initial metallicity) of
light elements and Fe-peak elements with those of weak r-process
elements as ``primary component''. In this case, the primary process
abundances include the abundances of the primary light elements,
the primary Fe-peak elements and the weak r-process elements.

The stellar chemical abundances reveal the contributions of
various nucleosynthesis processes and could not be explained by
only one process \citep{allen2006}.
Therefore the decomposition of the total abundance of each
observed element in a star is meaningful to research the relative
contribution of individual nucleosynthesis process.
One of our goals is to investigate the astrophysical origins of
the elements, especially the lighter n-capture elements in ET0097.
For this purpose, having adopted the abundance decomposition approach
presented by \citet{li2013a}, we explore the astrophysical origins
of the elements of this sample star.
Because the contribution of the main s-process is effective when
[Fe/H] $\geq$ $-$1.5 and the weak s-process is deemed as a secondary
process (i.e., the yields increase with increasing metallicity)
\citep{travaglio1999,travaglio2004}, in the first step, we neglect
the contributions of the s-process to ET0097. The $i$th element
abundance of ET0097 could be expressed as follows:
\begin{equation}
\label{abunequa1}
N_{i}=(C_{r,m}N_{i,r,m}+C_{pri}N_{i,pri})\times10^{[Fe/H]},
\end{equation}
where $N_{i,r,m}$ and $N_{i,pri}$ are the main $r$-process and
primary process abundances which are adopted from \citet{li2013a}
(per H = 10$^{12}$ at [Fe/H] = [Fe/H]$_{\odot}$).
$C_{r,m}$ and $C_{pri}$ are the corresponding component coefficients.

We obtain the best component coefficients by seeking for the
minimum $\chi^{2}$. The reduced $\chi^{2}$ is defined as follows:
\begin{equation}\label{fitequa1}
\chi^{2}=\sum_{i=1}^{K}
\frac{(\log N_{i,obs}- \log N_{i,cal})^{2}}
{(\Delta \log N_{i,obs})^{2}(K-K_{free})},
\end{equation}
where $\log\,N_{i,obs}$ is the observed abundance of the $i$th
element, $\Delta{\log N_{i,obs}}$ is the observed error, $N_{i,cal}$
is the calculated abundance, K is the number of the elements used by the
corresponding fit, and $K_{free}$ is the number of the free parameters.
With the best component coefficients, we can derive the calculated
abundances of ET0097. If each observed elemental abundance can
fit into the corresponding calculated abundance within the observed
error, the relative contribution of individual process to the elemental
abundance of the star can then be determined.

\section{Results and discussions}

\begin{figure*}
\centering
\includegraphics[width=12cm]{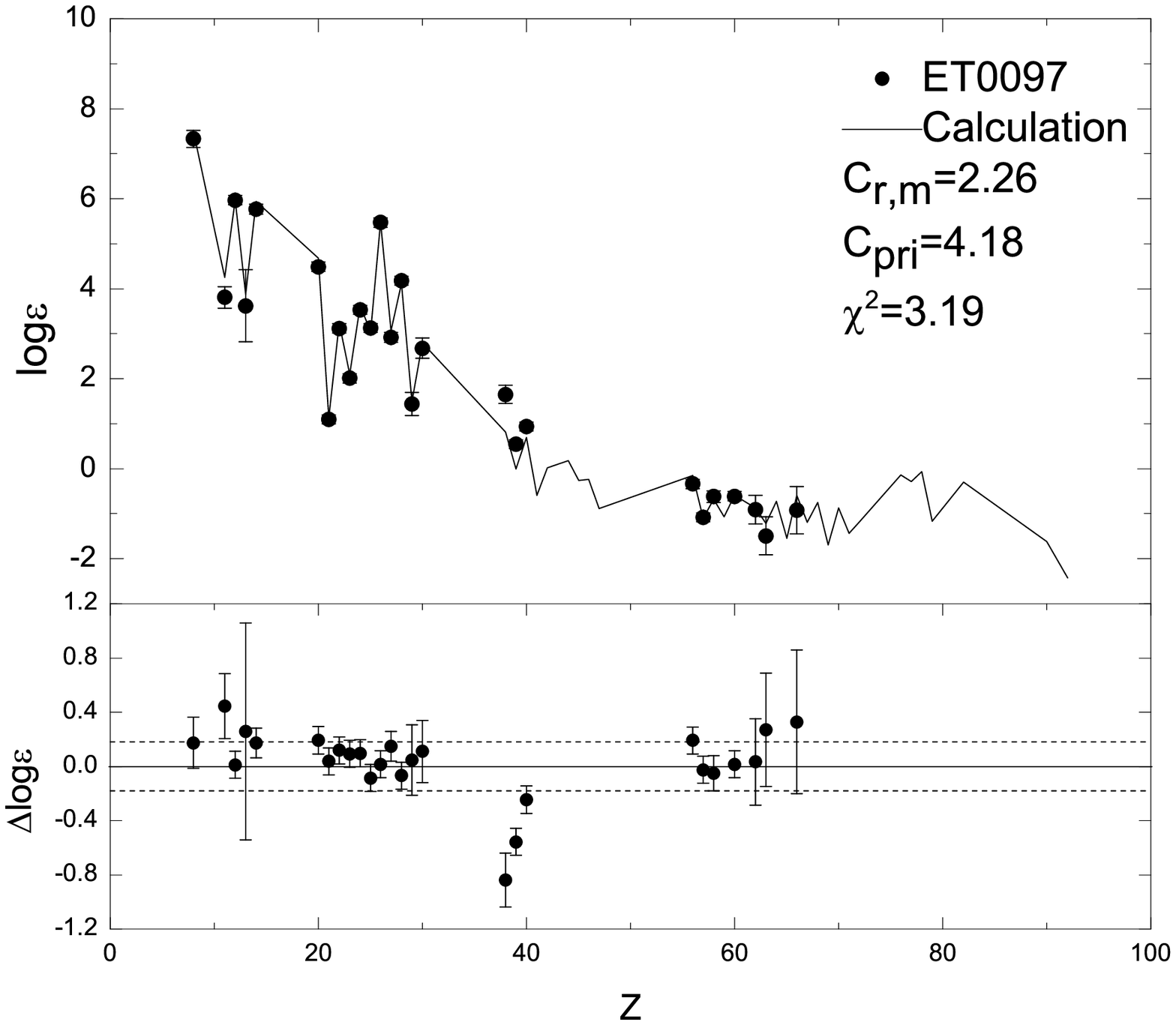}
\caption {Best fitted results in
the first step for the sample star. In
the upper panel, the filled circles are the observed abundances and
the solid line is the calculated abundances. In the lower panel, the
filled circles refer to the relative offsets, $\Delta
\log\,\varepsilon(x)\equiv \log\,\varepsilon(x)_{cal} -
\log\,\varepsilon(x)_{obs}$. The dashed lines are the calculated
errors. \label{fiteps1}}
\end{figure*}

In this work, we aim to determine the astrophysical origins
of the elements, especially the lighter n-capture elements in ET0097
and explore the relative contribution of individual nucleosynthesis
process to the elements of this CEMP star.
Based on equations (\ref{abunequa1}), (\ref{fitequa1}) and
the observed abundances adopted from \citet{skuladottir2015},
we start to explore the origins of the elements for the metal-poor
star ET0097 ([Fe/H] = $-$2.03) with the main r-process and primary process.
The best fitted results are shown in Figure \ref{fiteps1}.
In the upper panel, the observed abundances are plotted by the filled circles
and the calculated abundances are indicated by the solid line.
The lower panel shows the relative offsets, $\Delta
\log\,\varepsilon(x)\equiv \log\,\varepsilon(x)_{cal} -
\log\,\varepsilon(x)_{obs}$. Because the component coefficients are
constrained by the observed abundances, the calculated errors are
estimated from the average observed errors ($\simeq$ 0.18 dex) of
ET0097, which are shown by the dashed lines.
From Figure \ref{fiteps1}, we can see that, for most light
elements, Fe-peak elements and heavier n-capture elements, the
observed abundances can fit into the calculated abundances within
the observed errors, whereas the calculated abundances of the
lighter n-capture elements are lower than the corresponding observed
abundances. The discrepancy between the calculated abundances and the
observed abundances means that the contributions of the main r-process
and primary (or weak r-) process are not sufficient to explain the
abundances of the lighter n-capture elements of ET0097 and the
contribution of another process is required. Note that the
overabundances of lighter n-capture elements should not be ascribed
to the main s-process, because of the low observed ratio
[Ba/Fe] = $-$0.44 in ET0097 \citep{skuladottir2015}.

Historically, because of the secondary-like nature of the neutron source
$^{22}$Ne($\alpha,n$)$^{25}$Mg, the contribution of the weak s-process
of normal massive stars to metal-poor stars is neglected.
However, considering the rapid rotation of massive stars,
\citet{frischknecht2012} found that low-metallicity massive
stars should produce primary weak s-process elements
effectively.
Following this nucleosynthetic theory, \citet{cescutti2013} suggested
that metal-poor stars with high [Sr/Ba] should contain the s-process
material from rapidly rotating massive stars.
Recently, \citet{jablonka2015} carried out a detailed analysis of
the elemental abundances for five extremely metal-poor stars in Sculptor.
They found that the contributions of the main r- and weak r-process
are not enough to explain the observed abundances of the metal-poor star
ET0381 ([Fe/H] = $-$2.44), because of the high ratio [Sr/Ba] = 0.36 of
this star.
They considered that the abundances of some n-capture elements in ET0381
should have the contribution of the s-process of massive stars.
The results of these works imply that the contribution of the primary
weak s-process to the abundances of lighter n-capture elements
of some metal-poor stars, especially the metal-poor stars with high
[Sr/Ba] such as ET0097 ([Sr/Ba] = 1.15), should not be neglected.

Adding the contribution of the weak s-process, we use the combined
abundances of the main r-process, weak r-process and weak s-process
to fit the observed abundances of ET0097:
\begin{equation}
\label{abunequa2}
N_{i}=(C_{r,m}N_{i,r,m}+C_{pri}N_{i,pri}+C_{s,w}N_{i,s,w})\times10^{[Fe/H]},
\end{equation}
where $N_{i,s,w}$ is the primary weak s-process abundance which has
been normalized to the weak s-process abundance of Sr presented by
\citet{raiteri1993}. $C_{s,w}$ is the corresponding component
coefficient. The abundance $N_{i,s,w}$ is adopted from
\citet{frischknecht2012} (model B3, $\upsilon_{ini}/\upsilon_{crit}$
= 0.5), as the calculated abundance ratio [Sr/Ba] = 0.94 is close to
the observed abundance ratio [Sr/Ba] = 1.15 of ET0097. The
component coefficients $C_{r,m}$, $C_{pri}$ and $C_{s,w}$ could
reveal the relative contributions from the main r-, primary and weak
s-process, respectively, so the relative contribution of individual
process to the observed abundances can be derived by applying these
component coefficients.

\begin{figure*}
\centering
\includegraphics[width=12cm]{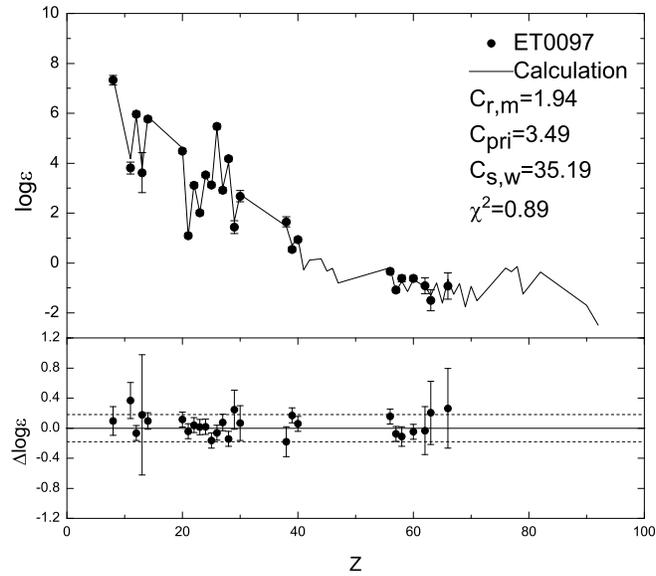}
\caption {New fitted results for the sample star.
Symbols are the same as in Figure \ref{fiteps1}.
\label{fiteps2}}
\end{figure*}

Considering the contribution from the primary weak s-process,
we investigate the origins of the elements for ET0097
using equations (\ref{fitequa1}) and (\ref{abunequa2}).
The new fitted results are shown in Figure \ref{fiteps2}.
The symbols are the same as in Figure \ref{fiteps1}. It is obvious
that all the calculated values are in excellent agreement with
the observed abundances within the observed errors. The results mean
that, although the contributions of the main r- and weak r-process
to the abundances of lighter n-capture elements are significant,
the observed overabundances of lighter n-capture elements of ET0097
could be ascribed to the contribution of the primary weak s-process of
massive stars, i.e., the CEMP star ET0097 should have the footprints
of the weak s-process.

\begin{figure*}
\centering
\includegraphics[width=12cm]{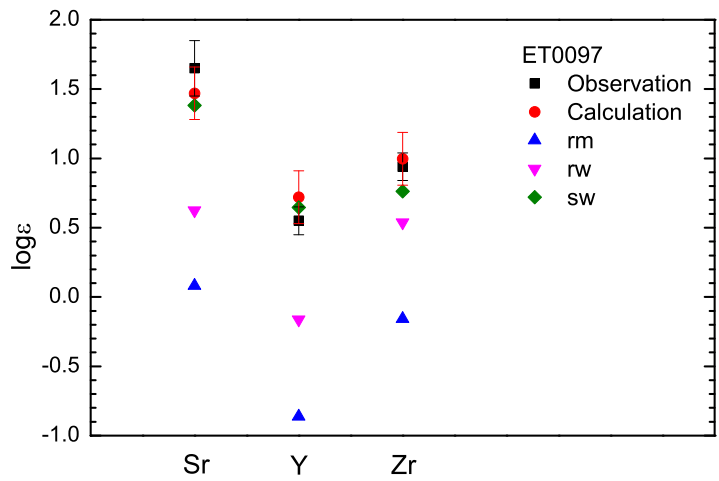}
\caption {Component abundances of the individual process for Sr, Y and Zr of the
sample star. The filled squares with error bars refer to the observed abundances
and the corresponding errors. The filled circles with error bars refer
to the calculated abundances and the corresponding errors.
The filled up triangles, filled down triangles and filled diamonds refer to
the abundances of the main r-process, weak r-process and primary weak s-process,
respectively. See the electronic edition of the Journal for a color version of this figure.
\label{sryzr}}
\end{figure*}

The observed abundances of Sr, Y and Zr can
fit successfully in the combined contributions of the main r-process,
the weak r-process and the primary weak s-process. Obviously, the relative
contributions of the three processes to the abundances of the lighter
n-capture elements are important for us to understand the elemental
origins of this star. The component abundances of the
three processes for Sr, Y and Zr of ET0097 are presented in
Figure \ref{sryzr}. The filled circles
with error bars refer to the calculated abundances and the
corresponding errors. The abundances of the main r-process, the weak
r-process and the primary weak s-process are plotted by the filled up
triangles, filled down triangles and filled diamonds, respectively.
From Figure \ref{sryzr}, we can see that for Sr, Y and Zr,
the abundances of the primary weak s-process are apparently higher than
those of the main r- and weak r-process, which means that the primary
weak s-process is the main contributor to the abundances of Sr, Y
and Zr of ET0097. The contributed fractions of the primary weak
s-process to Sr, Y and Zr abundances of ET0097 are about 82\%, 84\%
and 58\%, respectively. On the other hand, the contributed fractions
of the weak r-process to Sr, Y and Zr abundances are about 14\%, 13\%
and 35\%, respectively. Furthermore, the contributed fractions of
the main r-process to Sr, Y and Zr abundances are about 4\%, 3\% and
7\%, respectively. Obviously, the astrophysical reason of the
overabundances of the lighter n-capture elements in ET0097 can be
ascribed to the additional contribution of the primary weak
s-process. \citet{frischknecht2012} have predicted that
rapidly rotating massive stars with low metallicity should
facilitate the primary weak s-process. The derived footprints
of the weak s-process in ET0097 should be a significant evidence
that the weak s-process elements can be produced in metal-poor
massive stars.

For the CEMP-no star ET0097, the abundance ratios
[Sr, Y, Zr/Fe] = 0.47 ($\pm$ 0.11) and [Sr/Ba] = 1.15 ($\pm$ 0.22),
which means that the lighter n-capture elements are enhanced relative
to Fe and heavier n-capture elements. Based on the discussions above,
the astrophysical reason of enhanced lighter n-capture elements should
be the contributions of the primary weak s-process occurred in the rapidly
rotating massive stars with low metallicity. ET0097 should not be a
particular object and the similar abundance characteristics have also been
revealed in some CEMP-no stars of the Galactic halo. For the CEMP-no star
BS 16929-005, the observed abundance ratios [Sr, Y/Fe] = 0.22 ($\pm$ 0.16) and
[Sr/Ba] = 0.87 ($\pm$ 0.23) \citep{honda2004}. In addition, for the CEMP-no star
CS 22949-037, the abundance ratios [Sr, Y/Fe] = 0.15 ($\pm$ 0.24) and
[Sr/Ba] = 0.94 ($\pm$ 0.39) \citep{norris2001}. Overall, the number of
CEMP-no stars is still small, especially for the stars with
excessive abundances of lighter n-capture elements and deficient abundances of
heavier n-capture elements. Obviously, further abundance studies of
CEMP-no stars are needed.

\section{Conclusions}

The observations of n-capture elements for the metal-poor stars
provide an excellent chance to determine the abundance patterns
synthesized by various n-capture processes. ET0097 is the first
observed CEMP star in the Sculptor dwarf spheroidal galaxy.
The elemental abundances of this star
should contain significant nucleosynthetic information. In this
work, having adopted the abundance decomposition approach, we investigate
the astrophysical origins of the elements in ET0097. We find that
the light elements and the Fe-peak elements (from O to Zn) of this
star mainly originate from the primary process of massive stars
and the heavier n-capture elements (heavier than Ba)
mainly come from the main r-process. However, the lighter n-capture
elements (i.e., Sr, Y and Zr) should mainly come from the
primary weak s-process. The contributed fractions of the primary
weak s-process to the Sr, Y and Zr abundances of ET0097 are about 82\%,
84\% and 58\%, respectively.

Historically, the weak s-process contribution to metal-poor
stars is thought to be extremely small, due to the effect of
the secondary-like nature of the neutron source
$^{22}$Ne($\alpha,n$)$^{25}$Mg in massive stars \citep{travaglio2004}.
If this is the case, metal-poor ``weak s-process stars'' could
not be found. Recently, \citet{frischknecht2012} predicted that
rapidly rotating massive stars with low metallicity should
facilitate the primary weak s-process. Our calculated results mean
that the abundances of Sr, Y and Zr in ET0097 mainly come from the
weak s-process, i.e., the CEMP star ET0097 should have the
footprints of the weak s-process.
The derived result should be a significant evidence
that the weak s-process elements can be produced in metal-poor
massive stars. We wish the derived results in this work can provide
more information and more constraints on the n-capture nucleosynthesis
for low metallicity. Clearly, more observational data for
metal-poor stars, particularly for the ones in which lighter
n-capture elements are enhanced, would be important for
future works.

\acknowledgments

We thank the referee for constructive suggestions which
have improved this manuscript significantly.
This work has been supported by the National Natural Science Foundation
of China under Grants 11673007, 11273011, U1231119, 10973006, 11403007, 11547041
and 11273026, the Natural Science Foundation of Hebei Province under
Grant A2011205102, and the Program for Excellent Innovative
Talents in University of Hebei Province under Grant CPRC034.


\end{document}